\documentclass{mn2e}
\usepackage{epsf,times}
\usepackage{amsmath}
\newcommand{\etal}{{et al}\/.}
\begin{document}
\title[{\rm Herschel}-ATLAS: Radio-selected galaxies]{{\it Herschel}-ATLAS: far-infrared properties of
 radio-selected galaxies*}
\author[M.J.\ Hardcastle \etal]
{M.J.\ Hardcastle$^1$,
J.S.\ Virdee$^2$,
M.J.\ Jarvis$^1$,
D.G.\ Bonfield$^1$,
L. Dunne$^{3}$, 
S.\ Rawlings$^2$,
\newauthor
J.A.\ Stevens$^1$,
N.M. Christopher$^2$,
I.\ Heywood$^2$,
T. Mauch$^2$,
D.\ Rigopoulou$^{2,26}$,
\newauthor
A. Verma$^2$,
I.K.~Baldry$^5$, %LJMU = 5
S.P.~Bamford$^3$, % Notts
S.~Buttiglione$^6$,  % INAF = 6
A.~Cava$^8$, % IAC= 8
D.L.~Clements$^{7}$,
\newauthor
A. Cooray$^{23}$,
S.M.~Croom$^9$, % Sydney = 9
A.~Dariush$^4$, % Cardiff
G.~De Zotti$^{6,22}$,
S. Eales$^4$, % Cardiff = 2
J.~Fritz$^{11}$, % Gent = 11
\newauthor
D.T.~Hill$^{10}$,   % SUPA St Andrews
D. Hughes$^{25}$,
R.~Hopwood$^{20}$, 
E.~Ibar$^{13,19}$, % UK ATC, Edinburgh = 13
R.J.~Ivison$^{13}$, % UK ATC, Edinburgh
D.H.~Jones$^{12}$, % AAO Australia
\newauthor
J.~Loveday$^{15}$, % Sussex = 15
S.J. Maddox$^{3}$,
M.J. Micha{\l}owski$^{17}$,
M. Negrello$^{20}$, 
P.~Norberg$^{17}$, % SUPA, ROE = 17
\newauthor
M.~Pohlen$^4$, % Cardiff
M.~Prescott$^{5}$, % LJMU
E.E.~Rigby$^3$, % Nottingham
A.S.G.~Robotham$^{10}$,
G.~Rodighiero$^{21}$, %INAF Italy
\newauthor
D.~Scott$^{16}$,
R.~Sharp$^{12}$, % AAO Australia
D.J.B. Smith$^{3}$,
P. Temi$^{24}$, and
E.~van~Kampen$^{14}$ % ESO
\\
$^1$ School of Physics, Astronomy and Mathematics, University of
  Hertfordshire, College Lane, Hatfield AL10 9AB\\
$^2$ Oxford Astrophysics, Denys Wilkinson Building, University of Oxford, Keble Rd, Oxford
  OX1 3RH\\
$^{3}$ School of
 Physics \&\ Astronomy, University of Nottingham, Nottingham NG7 2RD\\
$^4$ School of Physics \&\ Astronomy, Cardiff University, The Parade, Cardiff, CF24 3AA\\
$^5$ Astrophysics Research Institute, Liverpool John Moores
University, Twelve Quays House, Egerton Wharf, Birkenhead, CH41 1LD\\
$^6$ INAF-Osservatorio Astronomico di Padova, Vicolo dell'Osservatorio 5, I-35122,
Padova, Italy\\
$^{7}$ Astrophysics Group, Blackett Lab, Imperial College London,
Prince Consort Road, London SW7 2AZ\\
$^8$ Instituto de Astrof\'isica de Canarias (IAC) and Departamento de
 Astrof\'isica de La Laguna (ULL), La Laguna, Tenerife, Spain\\
$^9$ Sydney Institute for Astronomy, School of Physics, University of
Sydney, NSW 2006, Australia\\
$^{10}$ SUPA, School of Physics and Astronomy, University of St.
Andrews, North Haugh, St. Andrews, KY16 9SS\\
$^{11}$ Sterrenkundig Observatorium, Universiteit Gent, Krijgslaan 281
 S9, B-9000 Gent, Belgium\\
$^{12}$ Anglo-Australian Observatory, PO Box 296, Epping, NSW 1710, Australia\\
$^{13}$ UK Astronomy Technology Centre, Royal Observatory, Edinburgh,
 EH9 3HJ\\
$^{14}$ European Southern Observatory, Karl-Schwarzschild-Strasse 2,
D-85748, Garching bei M\"unchen, Germany\\
$^{15}$ Astronomy Centre, University of Sussex, Falmer, Brighton, BN1 9QH\\
$^{16}$ Department of Physics and Astronomy, University of British Columbia, 6224 Agricultural Road, Vancouver, BC, V6T1Z1, Canada\\
$^{17}$ SUPA, Institute for Astronomy, University of Edinburgh, Royal
Observatory, Blackford Hill, Edinburgh EH9 3HJ\\
$^{18}$ Astrophysics Branch, NASA Ames Research Center, Mail Stop
2456,  Moffett Field, CA 94035, USA\\
$^{19}$ Institute for Astronomy, University of Edinburgh, Royal
Observatory, Blackford Hill, Edinburgh EH9 3HJ\\
$^{20}$ Department of Physics and Astronomy, The Open University,
Walton Hall, Milton Keynes, MK7 6AA\\
$^{21}$ Department of Astronomy, University of Padova, Vicolo
dell'Osservatorio 3, I-35122 Padova, Italy\\
$^{22}$ SISSA, Via Bonomea 265, I-34136 Trieste, Italy\\
$^{23}$ Department of Physics and Astronomy, University of California,
Irvine, CA 92697, USA\\
$^{24}$ Astrophysics Branch, NASA/Ames Research Center, MS 245-6, Moffett Field, CA 94035, USA\\
$^{25}$ Instituto Nacional de Astrof\'{i}sica, \'{O}ptica y Electr\'{o}nica, Aptdo. Postal 51 y 216, 72000, Puebla, Mexico\\
$^{26}$ Space Science \& Technology Department, Rutherford Appleton Laboratory,
Chilton, Didcot, Oxfordshire OX11 0QX
}
\maketitle
\vbox{\vskip -60pt}
\begin{abstract}
We use the {\it Herschel}-ATLAS science demonstration data to
investigate the star-formation properties of radio-selected galaxies
in the GAMA-9h field as a function of radio luminosity and redshift.
Radio selection at the lowest radio luminosities, as expected, selects
mostly starburst galaxies. At higher radio luminosities, where the
population is dominated by AGN, we find that some individual objects
are associated with high far-infrared luminosities. However, the
far-infrared properties of the radio-loud population are statistically
indistinguishable from those of a comparison population of radio-quiet
galaxies matched in redshift and K-band absolute magnitude. There is
thus no evidence that the host galaxies of these largely
low-luminosity (Fanaroff-Riley class I), and presumably
low-excitation, AGN, as a population, have particularly unusual
star-formation histories. Models in which the AGN activity in
higher-luminosity, high-excitation radio galaxies is triggered by
major mergers would predict a luminosity-dependent effect that is not
seen in our data (which only span a limited range in radio luminosity)
but which may well be detectable with the full {\it Herschel}-ATLAS
dataset.
\end{abstract}
\vskip -50pt
\begin{minipage}{17.5cm}
{\footnotesize * {\it Herschel} is an ESA space observatory with science instruments
provided by European-led Principal Invstigator consortia and with
important participation from NASA.}
\end{minipage}
\clearpage
\section{Introduction}
\label{intro}

It is well known that the increase in the star formation density of
the Universe with redshift (e.g. Madau \etal\ 1996) is paralleled by
an increase in the luminosity density of quasars (e.g. Boyle \&
Terlevich 1998), suggesting a link between galaxy assembly
and accretion onto massive black holes. It is much less obvious
whether this link is a direct one: is star-formation activity
physically associated with AGN activity, and, if so, with what types
of AGN activity is it associated? Although modelling suggests that the
link may be a simple causal one, in the sense that mergers can trigger
both AGN activity and star formation (e.g. Granato \etal\ 2004; di Matteo
\etal\ 2005) this is a question that must be settled
by direct observation of the AGN and star-formation properties of
large samples of galaxies. Far-infrared observations, probing dust
heated by young stars, provide one of the best methods of measuring
the star-formation rate, but in the past it has been hard to obtain
statistically robust samples, although this has improved recently
(e.g.\ Serjeant \& Hatziminaoglou 2009; Serjeant \etal\ 2010; Bonfield
\etal\ 2010).

Radio-loud active galaxies are an important population in the study of
the relationship between star formation and AGN activity. At the
lowest radio luminosities (a population which is best studied at $z
\sim 0$) the radio galaxy population is dominated by objects for which
there is no evidence at any waveband for any radiatively efficient AGN
activity, setting aside non-thermal emission associated with the
nuclear jet (e.g. Hardcastle \etal\ 2009 and references therein).
These objects have traditionally been called low-excitation radio
galaxies (e.g. Hine \& Longair 1979; Laing \etal\ 1994; Jackson \&
Rawlings 1997), but the differences between them and their
`high-excitation' counterparts are not only a matter of emission-line
strength, but extend to optical (Chiaberge \etal\ 2002), X-ray
(Hardcastle \etal\ 2006) and mid-IR (Ogle \etal\ 2006; Hardcastle
\etal\ 2009). The low-excitation objects, where the AGN power output
is primarily kinetic, clearly do not have the capability to regulate
and eventually terminate star formation through coupling of their
radiative output to cold gas. Instead the power of the AGN is put into
the expansion of radio lobes, and therefore predominantly does work on
the hot phase of the interstellar medium. At the highest radio
luminosities, though, radio-loud AGN activity is almost always
associated with radiatively efficient accretion, and these objects can
have strong effects on both the hot and cold gas in their environments
(cf. the `radio mode'/`quasar mode' dichotomy of Croton \etal\ 2006;
powerful radio-loud AGN are operating in both modes simultaneously).

The relationship between radio-loud AGN and star formation might thus
be expected to be a complex one, depending on radio luminosity,
redshift and possibly radiative efficiency of the AGN, and indeed
existing data present a picture of the relationship that seems to
depend strongly on what AGN population (or individual object) is
selected and which star-formation indicator is used. Hardcastle
\etal\ (2007) argued that the nuclear differences between low- and
high-excitation radio galaxies, suggesting a difference in accretion
mode, might be explained by different {\it sources} of the accreting
material, with the low-excitation sources powered by accretion (direct
or otherwise) of the hot phase of the IGM, while high-excitation
sources would be powered by cold gas supplied by mergers. This model
is quantitatively viable in specific cases (Hardcastle \etal\ 2007;
Balmaverde \etal\ 2008) and is qualitatively supported by the
observation that low-excitation sources of similar radio powers prefer
richer environments (Hardcastle 2004; Tasse \etal\ 2008). However, it
also makes a prediction that high-excitation sources will be
preferentially associated with gas-rich mergers and therefore star
formation, which is borne out by observations both at low redshift
(Baldi \& Capetti 2008) and at $z \sim 0.5$ (Herbert \etal\ 2010).

Far-infrared/sub-mm studies of star formation in samples of radio
galaxies have so far concentrated on high-redshift objects, in which
emission at long observed wavelengths (e.g. 850 $\mu$m, 1.2 mm)
corresponds to rest-frame wavelengths around the expected peak
  of thermal dust emission (e.g.\ Archibald \etal\ 2001; Reuland
\etal\ 2004). Working at these high redshifts with the available
flux-limited samples in the radio necessarily restricts these studies
to the most powerful radio-loud AGN. However, the availability of
observations at shorter far-IR wavelengths with the {\it Herschel
  Space Observatory} (Pilbratt \etal\ 2010) opens up the possibility
of studies of very large populations of more nearby objects. In this
paper, we present an analysis of all the radio-selected objects
identified with galaxies detected by {\it Herschel} in the 14 deg$^2$
field acquired in the Science Demonstration Phase (SDP) of the {\it
  Herschel} Astrophysical Terahertz Large Area Survey (H-ATLAS: Eales
\etal\ 2010). Although the full area of H-ATLAS (550 deg$^2$) will be
required to allow us to investigate the complete range of dependence
of host galaxy properties on radio luminosity and redshift, we show
that the new data shed some light on the nature of the hosts of the
numerically dominant low-luminosity radio galaxy population.

Throughout the paper we use a
concordance cosmology with $H_0 = 70$ km s$^{-1}$ Mpc$^{-1}$,
$\Omega_{\rm m} = 0.3$ and $\Omega_\Lambda = 0.7$.

\section{The data}
\label{data}

The following images and catalogues were available to us:

\begin{enumerate}

\item PSF-convolved, background-subtracted images of the SDP field at
  the wavelengths of 250, 350 and 500 $\mu$m provided by the Spectral
  and Photometric Imaging Receiver (SPIRE) instrument on {\it
    Herschel} (Griffin \etal\ 2010). The construction of these images
  for the H-ATLAS SDP data is described in detail by Pascale
  \etal\ (2010) and Rigby \etal\ (2010). We do not consider the PACS
  (Photodetector Array Camera and Spectrometer; see Ibar \etal\ 2010)
  data here as they are not deep enough for detections of more than a
  small fraction of our targets. The SDP field consists of two
  observations over a 16 deg$^2$ region of the sky centred at $\rm{RA}
  = 9^{\rm h} 5^{\rm m} 30^{\rm s}$, ${\rm Dec} = 0^\circ 5'
  0''$. We restricted our analysis to the sub-region of the
  SDP field in which there was good data from both {\it Herschel}
  scans (hereafter the `good' area of the SDP field); this covers
  14.38 deg$^{2}$.

\item A catalogue of FIR sources detected in the SDP field, which includes
  any source detected at $5\sigma$ or better at any SPIRE wavelength
  (Rigby \etal\ 2010).

\item Radio source catalogues and images from the NRAO VLA Sky Survey
  (NVSS) (Condon \etal\ 1998) and Faint Images of the Radio Sky at
  Twenty-one centimetres (FIRST: Becker, White \& Helfand 1995)
  surveys. These cover the whole of the SDP field.

\item Catalogues and images from the United Kingdom Infra-Red
  Telescope Deep Sky Survey -- Large Area Survey (UKIDSS-LAS,
  hereafter LAS: Lawrence \etal\ 2007). The LAS covers only 92 per
  cent of the area of the SDP field.

\item Redshifts from the Galaxy And Mass Assembly survey (GAMA: Driver
  \etal\ 2009, 2010). GAMA is a deep spectroscopic survey with a
  limiting depths of $r_{\rm AB} < 19.4$ mag, $z < 18.2$ and $K_{\rm
    AB} < 17.6$ in the SDP field; details of the target selection and
  priorities are given by Baldry \etal\ (2010) and Robotham
  \etal\ (2010). As described in more detail in Smith \etal\ (2010:
  hereafter S10), the GAMA catalogue for this area contains 12626 new
  spectroscopic redshifts in addition to 1673 redshifts from previous
  surveys in the area.

\item A catalogue of galaxies in the SDP field with photometric
  redshifts based on the LAS and Sloan Digital Sky Survey Data Release
  7 (SDSS-DR7: Abazajian \etal\ 2009) photometric data, as
  described by S10. We filter this catalogue so as to
  require a K-band detection, to exclude any sources that are
  point-like in either the LAS or SDSS parent catalogues, and to
  impose the magnitude selection $r<22$ mag.

\item A list of identifications between these galaxies and detected
  sources in the H-ATLAS data, again as described by S10. For these
  identifications S10 define a reliability $R$ which is a measure of
  whether a single optical ($r$-band) source dominates the observed
  FIR emission; they suggest that only sources with $R>0.8$ be used
  for this to be the case. Throughout the paper we consider all
  sources in the S10 catalogue, but distinguish in our analysis
  between `reliable' ($R>0.8$) and unreliable identifications.

\item A catalogue of optically selected quasars from the SDSS
  (Schneider \etal\ 2010) and 2dF-SDSS Luminous Red Galaxy and Quasar
  (2SLAQ: Croom \etal\ 2009) surveys in the SDP area, generated by
  Bonfield \etal\ (2010). We use this purely as a comparison
  population: the reader is referred to Bonfield \etal\ (2010) for
  details of the selection of these objects.

\end{enumerate}

\section{The radio-selected sample}
\label{sample}

We began by constructing a sample of radio-detected objects in the H-ATLAS
SDP field. This was done in the following way:

\begin{enumerate}
\item We selected all catalogued NVSS sources in the `good' area of
  the SDP field (796 in total). As the
  flux cutoff for the NVSS catalogue is $5\sigma$ (2.5 mJy), these are
  all clearly detected radio sources. The good short-baseline coverage
  of the NVSS
  data ensure that the NVSS flux densities are good estimates of the
  true total flux density of our targets.
\item We then cross-matched to the LAS K-band images by overlaying
  radio contours on LAS images, accepting only sources which had an
  association between the FIRST or, in a very few (5) cases, NVSS
  radio images and a K-band object with the appearance of a galaxy or
  a quasar. FIRST is used in preference to NVSS for identifications because of its much
  higher angular resolution, allowing less ambiguous identifications
  where compact radio components are present. This process excludes
  some weak or diffuse NVSS sources where FIRST detections were not
  available and where the NVSS position is inadequate to allow an
  identification with a LAS source. In addition, where NVSS sources
  were found to be blends of two or more FIRST sources, we corrected
  the NVSS flux density by scaling it by the ratio of the FIRST flux
  of the nearest source to the total FIRST flux. As mentioned above,
  the LAS covers only 92\% of the area of the SDP field, so the choice
  to use this as our reference catalogue slightly reduces our coverage
  but does not affect the sample completeness in any way. We have 391
  objects with LAS identifications.
\item Finally, we cross-correlated with the catalogue of galaxies with
  photometric redshifts (S10) on the basis of the LAS association.
  This gave us a total of 187 radio-loud sources in the SDP field.
  Where a spectroscopic redshift was available from the GAMA catalogue
  (including pre-existing SDSS redshifts) we used that in preference
  to a photometric redshift in subsequent analysis. 94 of our sources
  (50 per cent) had spectroscopic redshifts determined in this way.
  The median error on the photometric redshifts for all objects is
  0.03 (see S10 for a discussion of the errors).
\end{enumerate}

This process gives us a catalogue which is flux-limited in the radio
(by virtue of the original selection from the NVSS) and also
magnitude-limited in the optical (we require a K-band identification
and also require $r<22$). In practice, this means that we have few
sources with $z>0.7$ and none with $z>0.85$. The catalogue is also
likely to be strongly biased against radio-loud quasars since we have
excluded sources that appear point-like from our galaxy catalogue.
This has the desirable effect that the measured luminosities will tend
not to be strongly affected by beaming and that any contamination of
the fluxes measured at {\it Herschel} wavelengths by non-thermal
emission might be expected to be limited (cf. the results of Hes
\etal\ 1995). Analysis of the full catalogue without the restriction
of identification with (relatively) bright optical galaxies will be
presented elsewhere (Virdee \etal\ in prep.): here we concentrate on
the implications of the {\it Herschel} properties of these sources for
the nature of radio-loud AGN activity.

\begin{figure}
\epsfxsize 8.5cm
\epsfbox{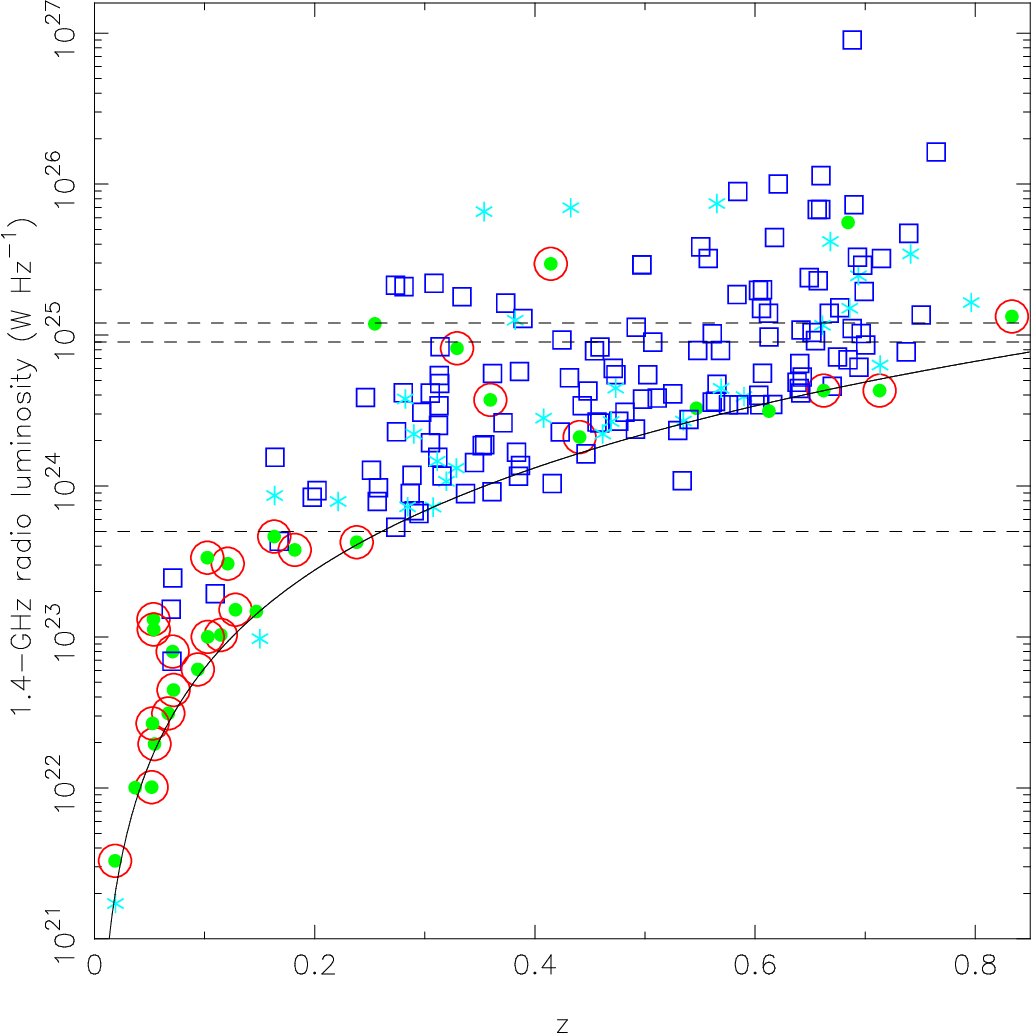}
\caption{Radio luminosity of the radio-loud sample as a function of
  redshift. Sources nominally detected by {\it Herschel} (i.e. sources
  identified with Herschel objects in the S10
  catalogue) are marked as filled green circles. If the association
  with the LAS galaxy is deemed `reliable' in the S10
  catalogue, the object is also marked with a red open circle.
  Any source not listed as detected in the catalogue, but detected
  down to the $2\sigma$ level in the 250-$\mu$m images, is shown as a
  light blue star. Non-detections are marked with blue open
  squares. The solid line corresponds to a nominal $5\sigma$ flux
  cutoff in the NVSS of 2.5 mJy. Sources lying significantly below
  this line do so as a result of the deblending process described in
  the text. Dashed horizontal lines show radio luminosities
  corresponding to (from bottom to top) the dividing line adopted in
  the text between starbursts and AGN; the expected luminosity for a
  maximal (4000 $M_\odot$ year$^{-1}$) starburst; and the luminosity
  corresponding to the Fanaroff-Riley break. }
\label{lz}
\end{figure}

Fig.\ \ref{lz} shows the radio luminosity-redshift plot for our
radio-loud sample (here and throughout the paper we adopt $\alpha =
0.8$, where $S \propto \nu^{-\alpha}$, for the $K$-correction in the
radio luminosity calculations; $\alpha = 0.8$ is a typical observed
value for low-frequency selected objects\footnote{For example, the
  mean 178-750 MHz spectral index for the 3CRR sample is 0.79; see
  http://3crr.extragalactic.info/ . At these redshifts the calculation
is insensitive to the exact value adopted.}, and we expect that the selection
against point-like optical objects will tend to select against
flat-spectrum radio sources). It will be seen that we probe a wide
range of radio luminosities. At the low-luminosity, low-redshift end,
we expect from existing analysis of the local 1.4-GHz luminosity
function (e.g. Mauch \& Sadler 2007) that the population will be
dominated by luminous star-forming galaxies rather than radio-loud
AGN, although a few AGN may still be present. The starburst luminosity
function cuts off steeply above a few $\times 10^{23}$ W Hz$^{-1}$ at
1.4 GHz, so we expect that almost all objects above this luminosity
will be radio-loud AGN, since we have no reason in the parent
catalogue to be biased towards starburst galaxies. The Fanaroff-Riley
break, i.e. the luminosity at which the population of radio galaxies
switches from being dominated by objects of Fanaroff \& Riley (1974)'s
morphological class I to mostly containing objects of Fanaroff-Riley
class II (hereafter these classes are abbreviated FRI and FRII) is at
$1.2 \times 10^{25}$ W Hz$^{-1}$ at 1.4 GHz, so our sample will be
numerically dominated by FRIs, i.e. objects that are traditionally
thought of as low-luminosity radio galaxies, but still has a
significant number of objects at FRII luminosities. We have made no
attempt to classify the objects morphologically in the radio using the
NVSS or FIRST data, as these datasets tend to lack the
surface-brightness sensitivity needed for reliable classification;
morphological investigations will be discussed in a future paper.

\section{The far-IR properties of the sample}

In this section we describe the properties of the sample in the
far-IR. Throughout this section, FIR fluxes in the SPIRE bands are
measured directly from the background-subtracted, PSF-convolved
H-ATLAS SDP images described in Section \ref{data}, taking the flux
density to be the value in the image at the pixel corresponding most
closely to the LAS position of our targets, with errors estimated from
the corresponding noise map. As discussed by Pascale \etal\ (2010),
PSF-convolved maps provide the maximum-likelihood estimator for the
flux density of an single isolated point source at a given position in
the presence of thermal noise; this remains a reasonable approximation
if there are small correlations between the positions of multiple
sources, as we expect in real data due to physical clustering of
objects. The flux densities we measure slightly underestimate the
total flux density if the source is resolved, as we have verified by
comparing our flux densities with those in the catalogue of Rigby
\etal\ (2010), but this is only likely to be a problem at the lowest
redshifts. We make an approximate correction for the mean background
due to confusion by subtracting the mean flux level of the whole
PSF-convolved map from each flux density measurement.

Throughout this section we distinguish between sources catalogued by
Rigby \etal\ (2010) and lower-significance detections, down to the
$2\sigma$ level, determined by the ratio of the flux density and the
error measured by us from the maps at 250 $\mu$m: this distinction is
only for the sake of illustration, in that it allows us to show the
properties of sources weaker than those in the source catalogue, and
the presence or absence of a $2\sigma$ detection forms no part of the
quantitative analysis in the paper.

To convert between measured flux or luminosity densities at {\it
  Herschel} SPIRE wavelengths and total flux or luminosity in the
far-IR band, as used in the literature, we have to adopt a model for
the far-IR spectral energy distribution (SED), since in general we do
not have enough data to fit models to each object. We investigated
several possible SED templates, including an optically thick grey-body
model whose parameters were determined from a fit to Arp 220 ($T =
61.7$K, $\beta = 1.54$, as used by Stevens \etal\ 2010), an optically
thin grey body with parameters fitted to M82 ($T = 44$ K, $\beta =
1.55$) and the model fitted to normal galaxies in the SDP field (the
S10 sample) by Dye \etal\ (2010) ($T = 26$ K, $\beta = 1.5$; cf.
  the very similar results, on a somewhat different sample of H-ATLAS
  sources, obtained by Amblard \etal\ 2010). These different
  choices of model parameters can make a significant difference (up
to an order of magnitude) to the inferred total fluxes or
luminosities, and also, via the $K$-correction, to their inferred
dependence on redshift. Because of this, we chose to use the model
adopted by Dye \etal , which has the merit of being derived from a
dataset that has considerable overlap with ours; for our current
  purposes the absolute normalization of the FIR luminosity is less
  important than the relative normalization, since our main
  conclusions will come from a comparison of the radio-loud objects
  with other samples. Detailed model fitting to the SEDs of
radio-loud objects, and thus the possibility of considering such
  factors as evolution in temperature with luminosity or redshift
will have to await a larger sample with a higher detection rate
and will probably require additional constraints, such as the
  use of the PACS data; we plan to address this in a future paper. For
  the present work, we integrate the Dye \etal\ model between 8 and
1000 $\mu$m to obtain the total FIR fluxes and luminosities, for
consistency with the approach used by other {\it Herschel} papers,
bearing in mind that this integration almost certainly underestimates
the total IR flux since it includes no component that radiates in the
mid-IR.

\subsection{Detections}
\label{detections}

\begin{figure}
\epsfxsize 8.5cm
\epsfbox{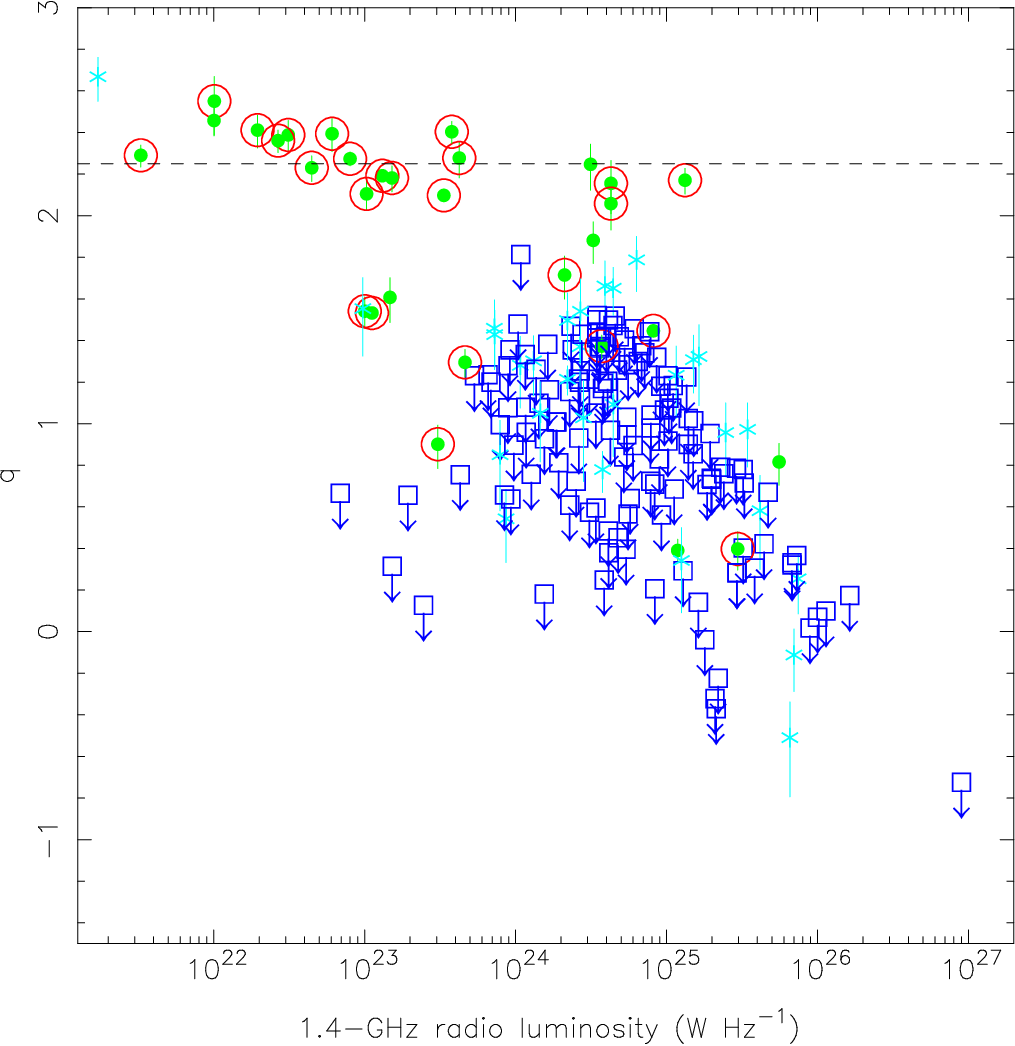}
\caption{The parameter $q$ (as defined in the text) as a function of
  radio luminosity for the sample. $q$ is estimated from the flux in
  the 250-$\mu$m maps in all cases as described in the text. Colours
  and symbols are as for Fig.\ \ref{lz}. The dashed horizontal line
  gives the median $q$ (2.25) for the 20 reliably detected sources
  with $L_{1.4} < 5 \times 10^{23}$ W Hz$^{-1}$. A $2\sigma$ upper
  limit on $q$ is plotted for undetected sources. }
\label{ql}
\end{figure}

As can be seen from Fig.\ \ref{lz}, only 31 of the galaxies identified
with our 187 sample objects have identifications with {\it Herschel}
sources from the catalogue of S10; all but 6 of these are classed as
`reliable' identifications, as discussed in Section \ref{data}. The
detection fraction is high at low radio luminosities (20/26 sources
below $L_{1.4} = 5 \times 10^{23}$ W Hz$^{-1}$ are detected) but then
drops off rapidly with increasing radio luminosity. Given the known
behaviour of the starburst and AGN luminosity functions (Section 2)
this is best explained in terms of a dominant starburst population at
low radio luminosities in our radio-selected sample. To illustrate
this, we have computed the parameter $q$, as originally defined by
Helou \etal\ (1985), for each of the detected objects. $q =
\log_{10}(S_{\rm IR}/(3.75 \times 10^{12} \times S_{1.4})$, where
$S_{\rm IR}$ is the integrated far-IR flux in W m$^{-2}$ and $S_{\rm
  1.4}$ is in W Hz$^{-1}$ m$^{-2}$. We estimate $q$ from the
250-$\mu$m flux density using the grey-body model discussed above,
taking proper account of $K$-correction. $q$ is plotted against radio
luminosity in Fig.\ \ref{ql}. For star-forming galaxies, $q$ by this
definition is expected to have a typical value around 2.4 independent
of radio luminosity, as found for example using {\it Herschel} data by
Ivison \etal\ (2010) and Jarvis \etal\ (2010), as a result of the
well-known radio-FIR correlation (e.g. Condon \etal\ 1991). We observe
that most of the low-luminosity sources have $q$ values in the range
2--3. However, most of the detected high-luminosity sources have
values of $q$ in the range 0-1.5, implying radio flux densities 1-2
orders of magnitude above the values expected from their FIR fluxes on
the radio-FIR correlation, and the upper limits for non-detections are
at a comparable level. We conclude that we are indeed seeing
predominantly starbursts at $L_{1.4} < 5 \times 10^{23}$ W Hz$^{-1}$
but that above this we are predominantly detecting genuine radio-loud
AGN. These results are unaltered if we use values of $q$ determined
from fits to the 70-500 $\mu$m SEDs of the detected {\it Herschel}
sources (Jarvis \etal\ 2010) and the values of $q$ obtained are also
consistent with that more detailed analysis. In what follows we use
$L_{1.4} < 5 \times 10^{23}$ W Hz$^{-1}$ as an approximate luminosity
cutoff to exclude the bulk of objects with $q$ values consistent with
being standard starbursts. While we recognise that more luminous
starbursts than this can and do exist, the small number of objects
observed to have $L_{1.4} > 5 \times 10^{23}$ W Hz$^{-1}$ and $q > 2$
suggests that few are present in our sample.

\subsection{Stacking analysis}
\label{stacks}

\begin{table*}
\caption{Mean bin flux densities and K-S probabilities that the {\it Herschel} fluxes of objects
  in redshift bins are drawn from the background distribution, as a
  function of wavelength. Low probabilities (below 1 per cent) imply
  significant differences between the bin being considered and the
  distribution of flux densities measured from randomly selected
  positions in the sky, as described in the text.}
\label{ks-z}
\begin{tabular}{llrrrrrrr}
\hline
Catalogued&$z$ range&Objects&\multicolumn{3}{c}{Mean bin flux density (mJy)}&\multicolumn{3}{c}{K-S probability (\%)}\\
sources?&&in bin&250 $\mu$m&350 $\mu$m&500 $\mu$m&250 $\mu$m&350 $\mu$m&500 $\mu$m\\
\hline
Included & 0.00 -- 0.10  &  15  & $396.1 \pm 1.7$  & $160.6 \pm 1.9$  & $58.8 \pm 2.3$ & $<10^{-3}$ & $<10^{-3}$ & $<10^{-3}$ \\
& 0.10 -- 0.20  &  14  & $125.1 \pm 1.7$  & $52.2 \pm 1.9$  & $18.0 \pm 2.4$ & $<10^{-3}$ & 0.004 & 0.7 \\
& 0.20 -- 0.35  &  39  & $13.3 \pm 1.0$  & $7.2 \pm 1.2$  & $2.7 \pm 1.4$ & 2.4 & 2.6 & 20.7 \\
& 0.35 -- 0.50  &  40  & $6.9 \pm 1.0$  & $5.6 \pm 1.1$  & $2.2 \pm 1.4$ & 0.07 & 2.3 & 40.1 \\
& 0.50 -- 0.65  &  43  & $5.9 \pm 1.0$  & $4.7 \pm 1.1$  & $3.6 \pm 1.4$ & 1.0 & 1.3 & 1.9 \\
& 0.65 -- 0.85  &  36  & $10.0 \pm 1.1$  & $5.5 \pm 1.2$  & $4.3 \pm 1.5$ & 0.007 & 3.9 & 5.2 \\
\hline
Excluded & 0.00 -- 0.10  &   4  & $133.7 \pm 3.3$  & $69.2 \pm 3.6$  & $26.3 \pm 4.5$ & 30.6 & 49.1 & 92.9 \\
& 0.10 -- 0.20  &   6  & $5.3 \pm 2.6$  & $6.0 \pm 3.0$  & $1.9 \pm 3.7$ & 14.9 & 25.7 & 33.1 \\
& 0.20 -- 0.35  &  36  & $3.1 \pm 1.1$  & $3.6 \pm 1.2$  & $1.9 \pm 1.5$ & 14.9 & 18.8 & 30.5 \\
& 0.35 -- 0.50  &  37  & $4.1 \pm 1.1$  & $4.1 \pm 1.2$  & $2.0 \pm 1.5$ & 0.5 & 4.5 & 59.0 \\
& 0.50 -- 0.65  &  41  & $3.6 \pm 1.0$  & $3.0 \pm 1.1$  & $2.9 \pm 1.4$ & 2.3 & 2.7 & 4.0 \\
& 0.65 -- 0.85  &  32  & $5.3 \pm 1.1$  & $2.1 \pm 1.3$  & $3.2 \pm 1.6$ & 0.05 & 19.0 & 9.0 \\
\hline
\end{tabular}
\end{table*}

\begin{table*}
\caption{Mean bin flux densities and K-S probabilities that the {\it Herschel} fluxes of objects in luminosity bins
  are drawn from the background distribution, as a function of
  wavelength. Notes as for Table \ref{ks-z}.}
\label{ks-l}
\begin{tabular}{llrrrrrrr}
\hline
Catalogued&Range in&Objects&\multicolumn{3}{c}{Mean bin flux density (mJy)}&\multicolumn{3}{c}{K-S probability (\%)}\\
sources?&$\log_{10}(L_{1.4})$&in bin&250 $\mu$m&350 $\mu$m&500 $\mu$m&250 $\mu$m&350 $\mu$m&500 $\mu$m\\
\hline
Included  & 21.0 -- 23.7 &  27  & $290.6 \pm 1.2$  & $118.0 \pm 1.4$  & $42.2 \pm 1.7$ & $<10^{-3}$ & $<10^{-3}$ & $<10^{-3}$ \\
 & 23.7 -- 24.3 &  32  & $2.4 \pm 1.1$  & $0.5 \pm 1.3$  & $-1.3 \pm 1.6$ & 28.3 & 69.8 & 12.5 \\
 & 24.3 -- 24.6 &  35  & $11.5 \pm 1.1$  & $10.7 \pm 1.2$  & $5.6 \pm 1.5$ & 0.010 & 0.05 & 1.1 \\
 & 24.6 -- 25.0 &  40  & $9.2 \pm 1.0$  & $4.5 \pm 1.1$  & $2.3 \pm 1.4$ & 0.8 & 10.3 & 48.4 \\
 & 25.0 -- 25.6 &  38  & $7.1 \pm 1.1$  & $4.7 \pm 1.2$  & $4.0 \pm 1.5$ & 0.9 & 12.2 & 5.1 \\
 & 25.6 -- 27.3 &  15  & $9.1 \pm 1.7$  & $7.2 \pm 1.9$  & $6.8 \pm 2.3$ & 0.10 & 0.4 & 1.2 \\
\hline
Excluded  & 21.0 -- 23.7 &   7  & $79.2 \pm 2.5$  & $42.7 \pm 2.8$  & $15.1 \pm 3.4$ & 15.2 & 39.2 & 99.5 \\
 & 23.7 -- 24.3 &  32  & $2.4 \pm 1.1$  & $0.5 \pm 1.3$  & $-1.3 \pm 1.6$ & 28.3 & 69.8 & 12.5 \\
 & 24.3 -- 24.6 &  31  & $6.6 \pm 1.2$  & $7.6 \pm 1.3$  & $4.7 \pm 1.6$ & 0.07 & 0.3 & 10.8 \\
 & 24.6 -- 25.0 &  37  & $2.5 \pm 1.1$  & $1.8 \pm 1.2$  & $2.0 \pm 1.5$ & 5.0 & 16.1 & 42.9 \\
 & 25.0 -- 25.6 &  35  & $3.2 \pm 1.1$  & $2.4 \pm 1.2$  & $3.0 \pm 1.5$ & 3.2 & 32.6 & 14.1 \\
 & 25.6 -- 27.3 &  14  & $7.7 \pm 1.7$  & $5.9 \pm 1.9$  & $6.6 \pm 2.4$ & 0.3 & 0.8 & 1.8 \\
\hline
\end{tabular}
\end{table*}

\begin{table*}
\caption{Mean bin flux densities of the comparison galaxy population
    and K-S probabilities that the {\it Herschel} fluxes of objects in redshift bins
  are drawn from this galaxy population, as a function of
  wavelength.}
\label{ks-zbin}
\begin{center}
\begin{tabular}{llrrrrrrr}
\hline
$z$ range&Radio-loud&\multicolumn{3}{c}{Mean galaxy flux density (mJy)}&\multicolumn{3}{c}{K-S probability (\%)}&Comparison\\
&objects in bin&250 $\mu$m&350 $\mu$m&500 $\mu$m&250 $\mu$m&350 $\mu$m&500 $\mu$m&galaxies\\
\hline
0.00 -- 0.10  &  15  & $22.3 \pm 0.2$  & $11.0 \pm 0.3$  & $4.7 \pm 0.3$ & $<10^{-3}$ & $<10^{-3}$ & $<10^{-3}$ & 770 \\ 
0.10 -- 0.20  &  14  & $15.5 \pm 0.1$  & $7.4 \pm 0.1$  & $3.0 \pm 0.2$ & 1.4 & 2.0 & 4.6 & 2913 \\ 
0.20 -- 0.35  &  39  & $12.6 \pm 0.1$  & $6.4 \pm 0.1$  & $2.9 \pm 0.1$ & 7.3 & 4.7 & 6.5 & 5481 \\ 
0.35 -- 0.50  &  40  & $7.9 \pm 0.1$  & $4.5 \pm 0.1$  & $2.0 \pm 0.1$ & 92.5 & 65.6 & 98.7 & 14660 \\ 
0.50 -- 0.65  &  43  & $6.9 \pm 0.1$  & $4.4 \pm 0.1$  & $2.2 \pm 0.1$ & 41.9 & 97.5 & 35.1 & 11905 \\ 
0.65 -- 0.85  &  36  & $9.5 \pm 0.1$  & $6.5 \pm 0.1$  & $3.1 \pm 0.2$ & 90.6 & 58.7 & 80.9 & 2971 \\ 
\hline
\end{tabular}
\end{center}
\end{table*}

Since the vast majority of our radio sources are undetected at the $5\sigma$
limit of the H-ATLAS source catalogue (Rigby \etal\ 2010), we need to use
statistical methods to calculate the properties of the source population.
We elected to stack the sample in bins corresponding to redshift and
radio luminosity to investigate the way in which the far-IR properties
varied with those two parameters.

To establish quantitatively whether sources in the bins were
significantly detected, we measured flux densities from 100,000
randomly chosen positions in the field; a Kolmogorov-Smirnov (K-S)
test could then be used to see whether the sources from our sample
were consistent with being drawn from a population defined by the
random positions. Using a K-S test rather than simply considering the
calculated uncertainties on the measured fluxes allows us to account
for the non-Gaussian nature of the noise as a result of confusion. We
chose the boundaries of our luminosity bins in such a way that the
lowest-luminosity bin contained all sources with $L < 5 \times
10^{23}$ W Hz$^{-1}$ and we also placed a bin boundary at the nominal
Fanaroff-Riley transition luminosity, while otherwise trying to keep
roughly similar numbers of sources per bin. Results are given in
Tables \ref{ks-z} and \ref{ks-l}, which also show the mean flux
density at each wavelength for every bin. We see that the sources in
the lowest luminosity bin and the lowest two redshift bins are very
strongly distinguished from the random background population, as
expected from Fig.\ \ref{lz}; sources at intermediate radio luminosity
or redshift are not distinguished from the background, but then
several higher-luminosity or higher-$z$ bins are distinguished from
the background at significance levels ranging from 2 to $>3\sigma$, at
least at 250 $\mu$m. As expected (since the beam is larger and the
potential for confusion greater), the significance drops off with
increasing wavelength. Importantly, these results still hold, though
obviously with somewhat reduced significance, if we exclude the
formally detected sources (those with identifications in the
catalogue) from the analysis, as shown in the bottom halves of
Tables \ref{ks-z} and \ref{ks-l}. Setting aside the lowest-luminosity
and lowest-redshift bins ($z<0.2$, $L_{1.4} < 5 \times 10^{23}$ W
Hz$^{-1}$), where the exclusion of detections removes almost all the
sources, all the bins that are significantly detected with the
inclusion of the formally detected sources are still detected at 95
per cent confidence or better on the K-S test at 250 $\mu$m if those
sources are excluded, and many bins are more significantly detected
than that; so at least at high radio luminosities or redshifts we are
not simply seeing the effect of a very few far-IR-bright objects. We
emphasise, though, that it is the K-S statistics that include the
detected objects (left-hand side of these tables) that determine
whether a given bin is actually detected.

We were also able to compare with the properties of galaxies that were
not identified with radio sources. To do this we selected all objects
identified as galaxies in the photometric redshift catalogue of S10
that had a LAS detection and a determined spectroscopic (GAMA) or
photometric redshift, imposing the cutoff $z<0.85$, but were not
identified with radio sources and that lay on the `good' SDP field
area so that photometry in the {\it Herschel} bands was possible. This
gave 59,817 galaxies in total. The population of radio-selected
sources is significantly ($>99.5$ per cent confidence) different from
this general galaxy population as a whole on a K-S test at all SPIRE
wavebands. However, if we break both populations down by redshift,
using the same binning scheme as previously and imposing the
additional requirement that the comparison galaxies lie in the same
K-band absolute magnitude range as the radio-loud hosts (which
restricts us to 38,700 galaxies), we see that this effect is
completely dominated by the sources with $z<0.2$, i.e. the sources
that we identify as star-forming galaxies (Table \ref{ks-zbin}). The
higher-redshift objects, which we expect to be radio-loud AGN, are
indistinguishable, statistically, in their FIR flux distribution from
galaxies selected in a similar way but without radio counterparts,
and their mean flux densities at each wavelength are also very
  similar. These similarities are understandable, since we know that,
at least at the low radio luminosities we are considering, radio
galaxy hosts very often have the appearance of passively evolving
ellipticals with no evidence either for strong star formation or for
an obscured, radiatively efficient AGN, and the K-band selection and
magnitude cutoff of our comparison population will have the effect of
including many similar objects. We return to this point below.

\subsection{Far-IR luminosity and star formation}
\label{sfr}

We estimated far-IR luminosities in our bins using a modified version
of the method of Serjeant \& Hatziminaoglou (2009). The value of this
approach is that it aims to account for the {\it intrinsic} scatter in
the flux density of objects in bins as well as the scatter due to
(thermal and confusion) noise when the luminosities are averaged; the
reader is referred to Serjeant \& Hatziminaoglou (2009) for a detailed
description of the method. For each redshift or luminosity bin, we
characterised the intrinsic scatter in the measured flux densities
using a maximum-likelihood fit. Serjeant \& Hatziminaoglou fitted a
Gaussian to the flux density distribution of each bin, but as Gaussian
fits were rather poor to some of our bins, we elected to use a
lognormal distribution in flux density (which also has the merit of
being constrained to positive flux density values), applying a prior
that is uniform in log space to the mean. Rather than assuming an
underlying Gaussian distribution of the noise, we used the actual
noise distribution in the data (as determined from our random flux
measurements) and used Monte-Carlo simulation to derive the
probability distribution of flux densities (which, after the addition
of noise, can be negative) and thus the model likelihood for any given
set of parameters. This allowed us to determine the best-fitting
values for the intrinsic distribution in flux densities using a
Markov-Chain Monte Carlo (MCMC) algorithm (see Mullin \& Hardcastle
2009 for a description of the code). Having done this, we took
noise-weighted means of the luminosities in each bin in the manner
described by Serjeant \& Hatziminaoglou (2009), for simplicity at this
stage making the approximation that both the intrinsic scatter in the
distribution and the noise were Gaussian so that they could be added
in quadrature to determine the weights. (We note that the overall
result of this procedure is reassuringly similar to what we obtain if
we simply take the mean of the luminosities using only the measured
noise values; the results are not dominated by the weights derived
from the intrinsic scatter estimates.) As in Section \ref{detections},
the luminosities were calculated from the measured 250-$\mu$m fluxes
(since the detections of stacks are most significant in this band, as
discussed in Section \ref{stacks}) on the assumption of a grey-body
template SED\footnote{The ratio of the stacked flux densities
    for these objects given in Table \ref{ks-z} allows us to make a
    rough check of the temperature of the grey-body model assumed in
    determining the luminosities. For temperatures in the range 20--30
    K and an assumed $\beta = 1.5$, the ratio of flux densities at 250
    and 350 $\mu$m, where we have the best statistics, should be in
    the range 1.9-2.7, with only a slight dependence on redshift. We
    see that this is broadly consistent with the flux densities in the
    stacks, although some have ratios closer to unity, which would
    imply even lower temperatures. However, given that we have no
    temperatures for individual sources and so must adopt a
    single-temperature model, it is reassuring that most of the flux
    density ratios are roughly consistent, within their uncertainties,
    with the $T=26$ K model that we are using.}.

Results are plotted in Fig.\ \ref{lum-stack}. Looking first at the
results of binning by radio luminosity (top-hand panel) we see that,
as expected, the lowest-luminosity radio bin has a reasonably high
mean FIR luminosity around $10^{11} L_\odot$. The highest FIR
luminosities of objects in this bin are comparable to those of
starbursts like Arp 220, which suggests a comparably high
star-formation rate. The second bin has a much lower luminosity
(which, given that this bin is not detected in our K-S test analysis,
should be considered to be an upper limit) corresponding to star
formation rates $\la5 M_\odot$ year$^{-1}$. This is the effect of
moving from a selection criterion that selects mostly starbursts to
one that selects mostly AGN, and it immediately shows that the
lowest-luminosity radio-loud AGN tend to have little or no star
formation. The remaining radio bins have FIR luminosities somewhat
higher than the mean in the first bin, and show at most a weak
positive trend with radio luminosity; these luminosities would
correspond (using the starburst relation from Kennicutt (1998), and
bearing in mind the many caveats associated with doing so) to total
star formation rates between 50 and 100 $M_\odot$ year$^{-1}$, which
might be associated either with the host galaxy of the radio source
itself (in which case these would be high star formation rates for
quiescent ellipticals) or with nearby galaxies in a host group or
cluster. It is important to bear in mind that the star-formation rates
we quote here, and the values plotted in Fig.\ \ref{lum-stack}, will
be affected by systematic uncertainties in the correction from
250-$\mu$m rest-frame flux to total mid-IR luminosity, as discussed at
the start of this Section. The {\it relative} star-formation rates
should be robust, but the absolute normalizations might be
systematically wrong by a significant factor. We note, however, that
the star-formation rates we derive are quite consistent with those
derived by Seymour \etal\ (2010) in a study of radio-loud objects in
the {\it Herschel} Multi-Tiered Extragalactic Survey (HerMES).

The results of luminosity stacking must be interpreted in the light of the stacking in
redshift bins (Fig.\ \ref{lum-stack}, lower panel). Here we perform an
identical stacking analysis using the normal galaxy population,
selected as defined in the previous Section, as a control. As we
expected given the results of the K-S tests, we see that the FIR
luminosities of radio-selected objects are much higher than those of
comparably selected galaxies in the lowest two redshift bins,
corresponding to the objects likely to be starburst galaxies. However,
above this redshift, we see very close agreement between the
radio-loud objects and the general galaxy population. The difference
between the two even at the highest redshifts (and therefore highest
radio luminosities) is little more than $1\sigma$ with the current data.

Finally, we can also compare with the properties of optically selected
quasars from the Bonfield \etal\ (2010) sample in the same redshift
bins. There are only 67 unique quasars in this sample in the $z<0.85$
redshift range, so the sample size is considerably smaller than for
our radio-selected objects; there are also no objects with $z<0.10$.
However, the individual redshift bins are all significantly detected
at 99 per cent confidence or better on K-S tests, so we can validly
use these small samples to compare with the radio galaxies. None of
the quasars in this redshift range is identified with a radio source
in our sample. We see (Fig. \ref{lum-stack}, lower panel) that there
is a clear trend for the luminosities of these quasars to lie
significantly above the luminosities of both the normal galaxies and
the radio-selected objects, though this tendency appears weaker at
higher redshift.

\section{Discussion and conclusions}

\begin{figure*}
\epsfxsize 13.8cm
\epsfbox{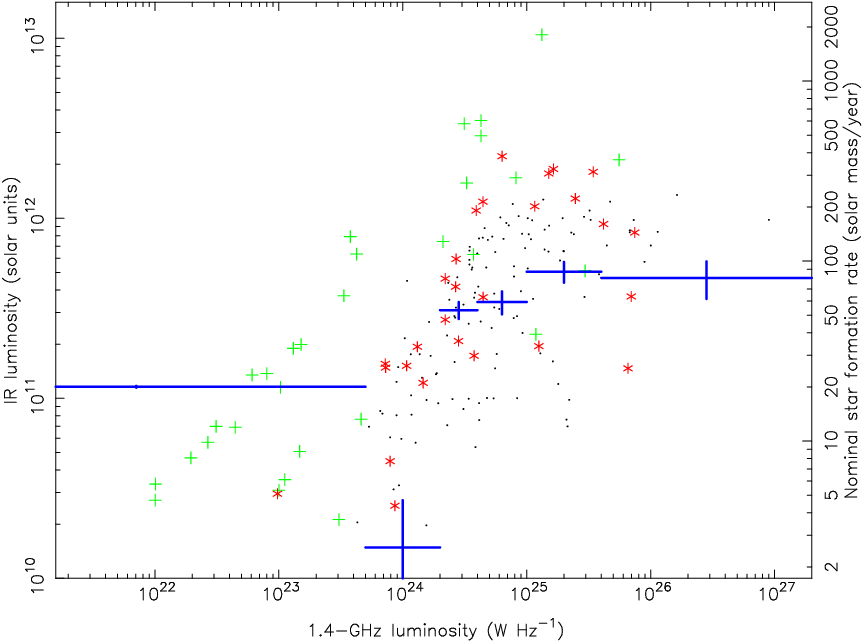}
\epsfxsize 13.8cm
\epsfbox{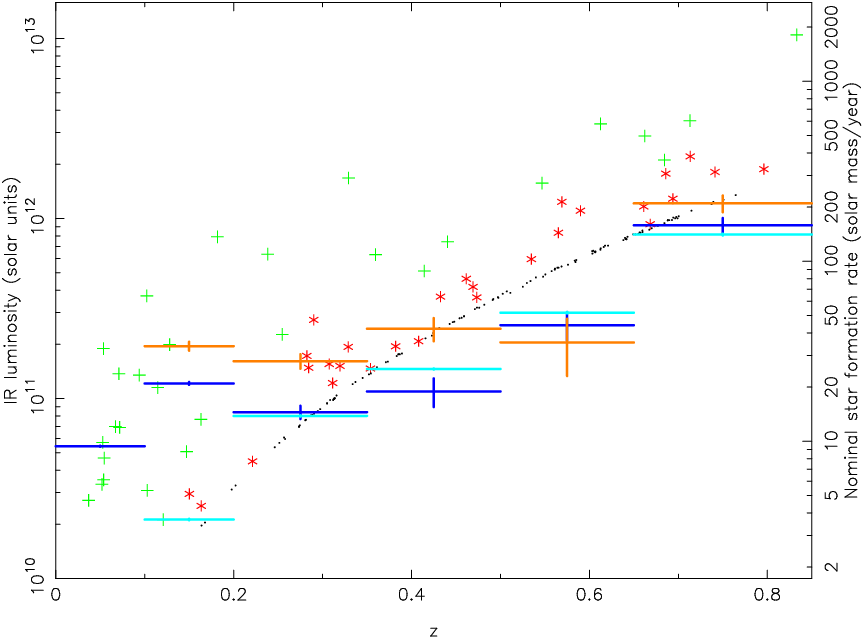}
\caption{Far-IR luminosity, derived from the measured 250-$\mu$m
  luminosity, as a function of (top) 1.4-GHz luminosity and (bottom)
  redshift for the radio-loud sample, showing both individual objects
  and the results of stacking as described in the text. Green crosses
  show identifications in the catalogue of S10; red stars show
  individual objects down to $2\sigma$, and black dots show $2\sigma$
  upper limits. Error bars on individual luminosities are not plotted
  for clarity. Blue solid lines in both panels are the results of
  stacking and show the mean luminosity and associated error bars in
  each of the radio luminosity or redshift bins defined in the text.
  Cyan solid lines in the lower panel show the results of a similar
  stacking analysis applied to the comparison sample of normal
  galaxies described in the text, and orange solid lines show the
  results of stacking the quasar sample also described in the text. The
  right-hand axis shows the conversion between far-IR luminosity and
  star-formation rate, assuming a Kennicutt (1998) starburst
  relationship; the locus of the limits in the bottom figure thus
  shows the limiting star-formation rate to which we are sensitive for
  individual objects.}
\label{lum-stack}
\end{figure*}

The most obvious conclusion to be drawn from Section \ref{sfr} is that
the FIR properties of radio galaxies and those of a
  comparably-selected population of radio-quiet objects are very
  similar, given the data that are presently available to us. The
selection of our comparison sample could clearly be improved with more
data; for example, our radio non-detections at high redshift will
clearly include some lower-luminosity radio-loud objects, while we
have not attempted to use the available optical data to select
exclusively elliptical galaxies. But taking the observations at face
value, we see no evidence that radio galaxy hosts behave any
differently in the FIR from a matched population of radio-quiet
galaxies. We do not believe that the incompleteness of our optical
identifications of the NVSS/FIRST sources should affect this
conclusion: the missing objects are likely either quasars (and
therefore undesirable because of the possibility of contamination by
non-thermal emission) or higher-redshift objects (see below).

Should we be surprised by our result? We begin by noting that we do
not expect any effect from any radiatively efficient AGN in these
radio galaxies --- the typical {\it mid-IR} luminosities (at around 15
$\mu$m) of radio-loud AGN of comparable radio luminosity to the most
powerful objects in our sample are 1-2 orders of magnitude lower than
what we observe in the present sample, if they are detected at all
(Hardcastle \etal\ 2009) and the emission from the torus of radio
galaxies believed to be hosting an obscured AGN would be expected, and
is observed, to peak in the mid-IR (e.g. Haas et al 2004). We also do
not expect to see any synchrotron contamination in the FIR, bearing in
mind that the typical flux density of our sources at 1.4 GHz is a few
mJy while the flux density at 250 $\mu$m (1.2 THz) of detected sources
is an order of magnitude higher, and that, as discussed above, our
selection criteria should strongly favour steep-spectrum objects. We
are thus safe to interpret the FIR luminosities as telling us about
star formation, subject to the usual caveats about young stars being
the dominant source of dust heating in these objects (Kennicutt 1998).

The picture appears then to be that the host galaxies of these
low-luminosity radio-loud AGN have, on average, no more -- or less --
star formation than the general population selected to have similar
K-band magnitudes. This is the first time it has been possible to
investigate this with a large sample in the FIR at these low radio
luminosities; earlier work studying high-$z$ radio galaxies and
radio-loud quasars with SCUBA (e.g. Archibald \etal 2001; Willott
\etal\ 2002; Rawlings \etal\ 2004) probed radio luminosities which
were almost all significantly larger than those studied here (for
example, all the sources in Archibald \etal\ have $L_{1.4} > 4 \times
10^{28}$ W Hz$^{-1}$), while studies of low-$z$ radio galaxies have
generally used shorter wavelengths, such as the 60 $\mu$m of {\it
  IRAS} (e.g. Yates \& Longair 1989; Hes \etal\ 1995) or the 70 $\mu$m
of {\it Spitzer} (e.g. Dicken \etal\ 2009) where the general consensus
seems to be that AGN-related thermal and non-thermal emission
dominates the measured flux densities. Our results will therefore be
useful for updating the latest models of far-infrared emission
emanating from AGN and their hosts (e.g. Wilman et al. 2010).

The result above is certainly consistent with our general picture of
the properties of radio-loud AGN and their hosts. We know that at
$z=0$ the vast majority of low-luminosity radio galaxies are hosted by
quiescent ellipticals (e.g., Lilly \& Longair 1984; Best \etal\ 2005);
in these low-power objects there is in general little evidence for
major, gas-rich mergers and so we would expect to see little or no
ongoing star formation (cf. Kaufmann, Heckman \& Best 2008). But it is
interesting to ask how this can be reconciled with the fact that there
does seem to be a strong association between AGN activity and star
formation when the AGN are selected at other wavebands (e.g. Serjeant
\etal\ 2010). The answer is probably related to the two populations of
radio-loud AGN, low-excitation and high-excitation radio galaxies,
described in Section \ref{intro}. The radio luminosity functions of
low-excitation and high-excitation radio galaxies clearly have
different slopes, since low-excitation objects are strongly
numerically dominant at low radio luminosities but almost completely
absent at the highest luminosities: the transition in terms of
numerical dominance seems to take place at radio luminosities
comparable to, but probably slightly above, the conventional
Fanaroff-Riley break. Our radio-loud sample is thus numerically
completely dominated by objects that are likely to be low-excitation
radio galaxies, which we would expect, based on the models discussed
in Section \ref{intro}, to have little or no association
with star formation. Only in the highest radio-luminosity bin might we
expect to see substantial numbers of high-excitation radio galaxies
with evidence for star formation, and the statistics there are poor.
It should be noted that Fig.\ \ref{lum-stack} clearly shows that there
are {\it individual} objects with both high radio luminosities and
high inferred star formation rates at all radio luminosities above a
few $\times 10^{24}$ W Hz$^{-1}$; these may well be high-excitation
radio galaxies associated with gas-rich mergers, and they certainly
account for the relatively high mean FIR luminosities and
star-formation rates in the high-radio-luminosity bins in that Figure.
The idea that high-excitation AGN, if present, would be associated
with higher star-formation rates is borne out, at least qualitatively,
by the position of the quasars from the sample of Bonfield
\etal\ (2010) in the lower panel of Fig.\ \ref{lum-stack}, though we
caution that we have necessarily made no attempt to match these in
host galaxy properties to the radio-selected or normal galaxies.

With deeper optical/IR data, better spectroscopy and radio data from
our GMRT observations, plus broader coverage with {\it Herschel} (all
of which will be available in the near future), we can test this
qualitative picture in a number of ways. With more objects of higher
radio luminosity, we will be able to break the luminosity-redshift
degeneracy inherent to the present sample (the results presented here
give no new information on whether the far-IR luminosity of radio
galaxy hosts depends principally on redshift or on radio luminosity, a
problem first identified by Rawlings \etal\ 2004). The SDP data
represent only 1/40th of the 570 deg$^2$ of the full H-ATLAS data, so
our eventual sample size will increase by a large factor: in
particular, we expect $\sim 6$ FRIIs per square degree (Wilman
\etal\ 2008) up to the highest redshift so the full H-ATLAS will give
us $>3000$ powerful sources. In addition, even with the current data
we have optical/IR identifications for less than half of the original
radio sample (Section \ref{sample}) so in principle many high-redshift
radio sources are already there and just await identification; the
availability of VIKING data (1.4 mag deeper than UKIDSS-LAS) will
allow us to identify powerful radio sources out to $z \sim 2.5$
(Jarvis \etal\ 2001; Willott \etal\ 2003) and so we will be able both
to extend the stacking study and, probably, to increase the number of
{\it Herschel}-detected sources, allowing a study of the detected
population as a function of radio luminosity and morphology. Secondly,
the model outlined above implies that we would expect any excess FIR
luminosity in radio galaxies to be associated not only with indicators
of star formation at other wavelengths (e.g. optical galaxy colours)
but also with indicators of high-excitation AGN activity such as
strong narrow emission lines (which can be investigated with GAMA and
other spectroscopy), a mid-IR excess (which might be visible in survey
data from the {\it Wide-field Infrared Survey Explorer}, {\it WISE})
and obscured nuclear X-ray emission (many H-ATLAS fields have
available X-ray data). With the larger sample and the wider
availability of multi-wavelength diagnostics provided by the full
H-ATLAS project, it should be possible to carry out a definitive test
of this model.

\section*{Acknowledgements}

{\it Herschel} is an ESA space observatory with science instruments
provided by European-led Principal Invstigator consortia and with
important participation from NASA. U.S. participants in {\it
  Herschel}-ATLAS acknowledge support provided by NASA through a
contract issued from JPL. GAMA is a joint European-Australian project,
based around a spectroscopic campaign using the AAOmega instrument,
and is funded by the STFC, the ARC, and the AAO. MJH thanks the Royal
Society for generous financial support through the University Research
Fellowships scheme. JSV thanks the STFC and RAL for a studentship. MJJ
acknowledges support from an RCUK fellowship. We thank an anonymous
referee for comments that have allowed us to improve the presentation
of the paper.

\clearpage

\begin{thebibliography}{}
\bibitem[]{0}Abazajian, K.N., et al. 2009, ApJS, 182, 543
\bibitem[]{30}Amblard, A., et al. 2010, A\&A, 518, L9
\bibitem[]{44}Archibald, E.N., Dunlop, J.S., Hughes, D.H., Rawlings, S., Eales, S.A., Ivison, R.J., 2001, MNRAS, 323, 417
\bibitem[]{62}Baldry, I.K., et al. 2010, MNRAS, 404, 86
\bibitem[]{65}Balmaverde, B., Baldi, R.D., Capetti, A., 2008, A\&A, 486, 119
\bibitem[]{101}Best, P.N., Kauffmann, G., Heckman, T.M., Brinchmann, J., Charlot, S., Ivezi\'c, Z., White, S.D.M., 2005, MNRAS, 362, 25
\bibitem[]{164}Boyle, B.J., Terlevich, R.J., 1998, MNRAS, 293, L49
\bibitem[]{256}Chiaberge, M., Macchetto, F.D., Sparks, W.B., Capetti, A., Allen, M.G., Martel, A.R., 2002, ApJ, 571, 247
\bibitem[]{285}Condon, J.J., Anderson, M.L., Helou, G., 1991, ApJ, 376, 95
\bibitem[]{286}Condon, J.J., Cotton, W.D., Greisen, E.W., Yin, Q.F., Perley, R.A., Taylor, G.B., Broderick, J.J., 1998, AJ, 115, 1693
\bibitem[]{316}Croom, S., et al. 2009, MNRAS, 392, 19
\bibitem[]{324}Croton, D., et al., 2006, MNRAS, 365, 111
\bibitem[]{349}di~Matteo T., Springel, V., Hernquist, L., 2005, Nat, 433, 604
\bibitem[]{350}Dicken, D., et al., 2009, ApJ, 694, 268
\bibitem[]{363}Driver, S.P., et al. 2009, A\&G 50 5.12
\bibitem[]{364}Driver, S.P., et al. 2010, MNRAS  submitted
\bibitem[]{368}Dye, S., et al. 2010, A\&A in press, arXiv:1005.2411
\bibitem[]{372}Eales, S., et al. 2010, PASP, 122, 499
\bibitem[]{413}Fanaroff, B.L., Riley, J.M., 1974, MNRAS, 167, 31P
\bibitem[]{510}Granato, G.L., De~Zotti, G., Silva, L., Bressan, A., Danese, L., 2004, ApJ, 600, 580
\bibitem[]{520}Griffin, M.J., et al. 2010, A\&A in press, arXiv:1005.5123
\bibitem[]{527}Haas, M., et al., 2004, A\&A, 424, 531
\bibitem[]{540}Hardcastle, M.J., 2004, A\&A, 414, 927
\bibitem[]{564}Hardcastle, M.J., Evans, D.A., Croston, J.H., 2006, MNRAS, 370, 1893
\bibitem[]{566}Hardcastle, M.J., Evans, D.A., Croston, J.H., 2007, MNRAS, 376, 1849
\bibitem[]{567}Hardcastle, M.J., Evans, D.A., Croston, J.H., 2009, MNRAS, 396, 1929
\bibitem[]{609}Helou, G., Soifer, B.T., Rowan-Robinson, M., 1985, ApJ, 298, L7
\bibitem[]{613}Herbert, P.D., Jarvis, M.J., Willott, C.J., McLure, R.J., Mitchell, E., Rawlings, S., Hill, G.J., Dunlop, J.S., 2010, MNRAS  in press
\bibitem[]{616}Hes, R., Barthel, P.D., Hoekstra, H., 1995, A\&A, 303, 8
\bibitem[]{623}Hine, R.G., Longair, M.S., 1979, MNRAS, 188, 111
\bibitem[]{661}Ivison, R., et al 2010, A\&A in press (arXiv:1005.1072)
\bibitem[]{670}Jackson, N., Rawlings, S., 1997, MNRAS, 286, 241
\bibitem[]{678}Jarvis, M.J., Rawlings, S., Eales, S., Blundell, K.M., Bunker, A.J., Croft, S., McLure, R.J., Willott, C.J., 2001, MNRAS, 326, 1585
\bibitem[]{679}Jarvis, M.J., et al. 2010, MNRAS submitted
\bibitem[]{721}Kauffmann, G., Heckman, T.M., Best, P.N., 2008, MNRAS, 384, 953
\bibitem[]{786}Laing, R.A., Jenkins, C.R., Wall, J.V., Unger, S.W., 1994, in Bicknell G.V., Dopita M.A., Quinn P.J., eds, The First Stromlo Symposium: the Physics of Active Galaxies, ASP Conference Series vol. 54, San Francisco, p.~201
\bibitem[]{832}Lilly, S.J., Longair, M.S., 1984, MNRAS, 211, 833
\bibitem[]{885}Madau, P., Ferguson, H.C., Dickinson, M.E., Giavalisco, M., Steidel, C.C., Fruchter, A., 1996, MNRAS, 283, 1388
\bibitem[]{917}Mauch, T., Sadler, E., 2007, MNRAS, 375, 931
\bibitem[]{964}Mullin, L.M., Hardcastle, M.J., 2009, MNRAS, 398, 1989
\bibitem[]{1001}Ogle, P., Whysong, D., Antonucci, R., 2006, ApJ, 647, 161
\bibitem[]{1082}Pilbratt, G.L., et al., 2010, A\&A in press (arXiv:1005.5331)
\bibitem[]{1109}Rawlings, S., Willott, C.J., Hill, G.J., Archibald, E.N., Dunlop, J.S., Hughes, D.H., 2004, MNRAS, 351, 676
\bibitem[]{1133}Robotham, A..., et al. 2010, PASA, 27, 76
\bibitem[]{1209}Schneider, D.P., et al 2010, AJ, 139, 2360
\bibitem[]{1216}Serjeant, S., Hatziminaoglou, E., 2009, MNRAS, 397, 265
\bibitem[]{1217}Serjeant, S., et al. 2010, A\&A in press (arXiv:1005.2410)
\bibitem[]{1235}Smith, D.J.B., et al 2010, MNRAS submitted [S10]
\bibitem[]{1308}Tasse, C., Best, P.N., R\"ottgering, H., Le~Borgne, D., 2008, A\&A, 490, 893
\bibitem[]{1399}Willott, C.J., Rawlings, S., Archibald, E.N., Dunlop, J.S., 2002, MNRAS, 331, 435
\bibitem[]{1400}Willott, C.J., Rawlings, S., Jarvis, M.J., Blundell, K., 2003, MNRAS, 339, 173
\bibitem[]{1402}Wilman, R.J., et al., 2008, MNRAS, 388, 1335
\bibitem[]{1439}Yates, M.G., Longair, M.S., 1989, MNRAS, 241, 29
\end{thebibliography}
\end{document}